\documentclass[lettersize,journal]{IEEEtran}
\usepackage{amsmath,amsfonts}
\usepackage{algorithmic}
\usepackage{array}
\usepackage[caption=false,font=normalsize,labelfont=sf,textfont=sf]{subfig}
\usepackage{textcomp}
\usepackage{stfloats}
\usepackage{url}
\usepackage{verbatim}
\usepackage{graphicx}
\usepackage[caption=false,font=normalsize,
   labelfont=sf,textfont=sf]{subfig}
\usepackage{cite}
\usepackage{bm}
\usepackage{hhline}
\usepackage[capitalise]{cleveref}
\usepackage{multirow}
\usepackage{array}
\usepackage{color}
\newcolumntype{P}[1]{>{\centering\arraybackslash}p{#1}}
\DeclareUnicodeCharacter{0301}{\'{e}}
\begin{document}

\title{Memory-Efficient FPGA Implementation of Stochastic Simulated Annealing}

\author{Duckgyu Shin, Naoya Onizawa, Warren J. Gross, Takahiro Hanyu}
\maketitle

\begin{abstract}
Simulated annealing (SA) is a well-known algorithm for solving combinatorial optimization problems.
However, the computation time of SA increases rapidly, as the size of the problem grows.
Recently, a stochastic simulated annealing (SSA) algorithm that converges faster than conventional SA has been reported.
In this paper, we present a hardware-aware SSA (HA-SSA) algorithm for memory-efficient FPGA implementations.
HA-SSA can reduce the memory usage of storing intermediate results while maintaining the computing speed of SSA.
For evaluation purposes, the proposed algorithm is compared with the conventional SSA and SA approaches on maximum cut combinatorial optimization problems.
HA-SSA achieves a convergence speed that is up to 114-times faster than that of the conventional SA algorithm depending on the maximum cut problem selected from the G-set which is a dataset of the maximum cut problems.
HA-SSA is implemented on a field-programmable gate array (FPGA) (Xilinx Kintex-7), and it achieves up to 6-times the memory efficiency of conventional SSA while maintaining high solution quality for optimization problems.
\end{abstract}
\begin{IEEEkeywords}
Stochastic computing, Simulated annealing, FPGA, Combinatorial optimization problem
\end{IEEEkeywords}
\section{Introduction}
\label{sec:intro}
Simulated annealing (SA) is a very effective probabilistic algorithm for finding the near-optimal solutions of combinatorial optimization problems \cite{SA}.
It can be applied to several practical applications such as mobile robot path planning \cite{path_finding} or very-large-scale-integrated (VLSI) circuit design \cite{sa_circuit_design}.
However, the computational time of SA increases rapidly as the size of the problem grows.
Non-conventional computing methods can be employed to accelerate SA, such as quantum annealing \cite{QA, QA2} or stochastic simulated annealing (SSA) \cite{tnnls}.
In \cite{tnnls}, a combinatorial optimization problem was converted to an Ising model \cite{ising_1} which is a network consisting of Ising spins and their interconnections.
The near-optimal solution to the optimization problem could be found through the convergence of the Ising energy (Hamiltonian) to a global minimum \cite{ising_sa}.
The Ising spin was realized in SSA by a probabilistic bit (p-bit) \cite{pbit_model} model using stochastic computing \cite{tnnls}.
The SSA algorithm in \cite{tnnls} is implemented in software and a small-scale hardware proof-of-concept was reported.
To solve practical large SA applications it is necessary to implement SSA hardware.
In SSA hardware, because of the stochastic nature of the algorithm, spin states need to be stored to determine the near-optimal solution.
This results in large usage which can limit the scalability of SSA hardware.
In this paper, we propose a hardware-aware SSA (HA-SSA) algorithm for implementing memory-efficient SSA hardware.
The HA-SSA algorithm reduces memory usage by deciding when to store a subset of spin states during processing.
Additionally, the HA-SSA algorithm also introduces a hardware-friendly pseudoinverse temperature control that has lower hardware costs.
For evaluation purposes, the HA-SSA algorithm is simulated and compared to the conventional SSA algorithm and the SA algorithm.
Maximum cut combinatorial optimization problems (MAX-CUT) \cite{maxcut} are solved by HA-SSA, SSA, and SA in simulation.
Moreover, we develop HA-SSA hardware implemented on a Xilinx Kintex-7 FPGA and evaluate the MAX-CUT on the implemented hardware.
The proposed HA-SSA hardware reduces the memory usage to 1/6 that of SSA while maintaining the same accuracy.
In addition, we compare the proposed hardware to an existing FPGA Ising-model solver with parallel tempering (PT) \cite{iccad_pt}.
For the same MAX-CUT problem, the proposed hardware achieves the best known solution with an annealing time that is approximately 2.6-times faster.
The contributions of this work are a hardware-friendly and memory-efficient optimization of the SSA algorithm and the implementation of large-scale annealing hardware.
The proposed annealing hardware can obtain the solution to the combinatorial optimization problem rapidly and offers a possible solution for adapting the SA algorithm to real-world problems.
The rest of this paper is organized as follows.
\cref{sec:pre} reviews the SA algorithm on the Ising model and the SSA algorithm.
\cref{sec:hassa} introduces the proposed HA-SSA algorithm.
\cref{sec:setup} describes the experimental conditions and the implementation of the proposed HA-SSA hardware on FPGA.
\cref{sec:eval} evaluates the HA-SSA algorithm and the annealing hardware and conducts comparisons with the methods in related work \cite{iccad_pt}.
Finally, conclusions are stated in \cref{sec:conc}.
\section{Preliminaries}
\label{sec:pre}
\subsection{SA on the Ising models}
\label{sec:sa}
\begin{figure}
    \centering
    \subfloat[]{
        \includegraphics[width=0.3\linewidth]{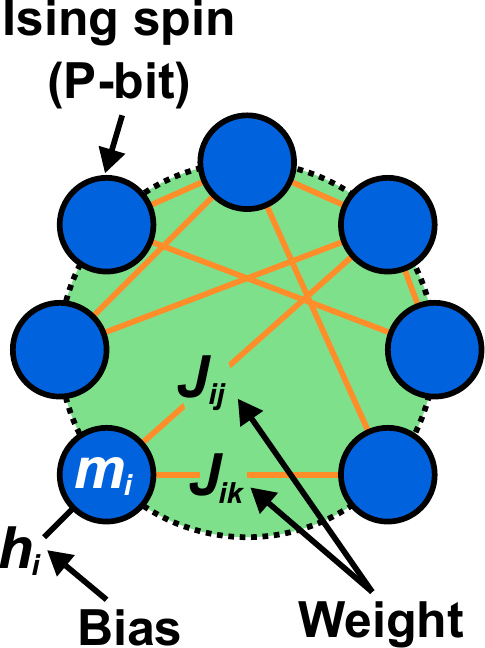}
        \label{fig:f1a}
    }
    \subfloat[]{
        \includegraphics[width=0.4\linewidth]{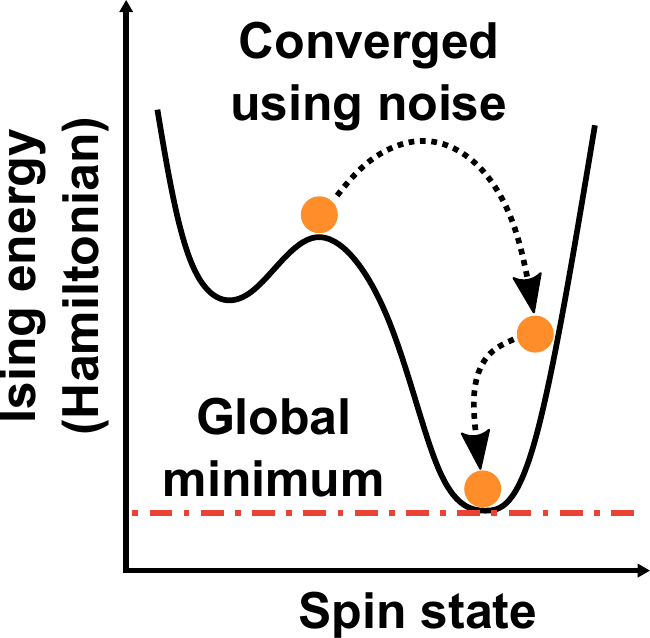}
        \label{fig:f1b}
    }
    \caption{The structure of the Ising model and its energy landscape. (a) The configuration of the Ising model for SA. (b) The transition of the spin state and the convergence of the Ising energy.}
    \label{fig:f1}
\end{figure}
A combinatorial optimization problem can be converted to an Ising model \cite{tnnls}.
First, the given optimization problem is converted to a quadratic unconstrained binary optimization (QUBO) model, because the QUBO model is equivalent to the Ising model \cite{qubo}.
The converted Ising model is represented as a network, as shown in \cref{fig:f1} (a), and it consists of Ising spins and their interconnections.
Each spin has a bias ($h_i$) and is connected to other spins with an interconnection weight of $J_{ij}$, and each spin state ($m_i$) has a value in $\{-1, +1\}$.
$h_i$ and $J_{ij}$ are determined while converting the problem to the Ising model so that the optimal solution of the problem is embedded at a global minimum (ground-state) of the Ising energy (Hamiltonian) \cite{20k_spin}.
The Hamiltonian ($H$) is given by:
\begin{equation}
    H = - \sum_{i} h_i m_i(t) - \frac{1}{2} \sum_{i, j} J_{ij} m_i(t) m_j(t).
    \label{eq:ham}
\end{equation}
\cref{fig:f1} (b) describes the convergence of the Ising energy.
During the annealing process, a randomly selected spin changes its state, and then the Ising energy is calculated.
When the new Ising energy is lower than the previous energy, the flipped state is accepted.
In contrast, when the new Ising energy is higher than the previous value, the acceptance of the new state is stochastically determined.
Since the flipped state is stochastically accepted, the Ising energy can escape from a local minimum and continue converging to the global minimum.
\subsection{SSA algorithm}
\label{sec:ssa}
In SSA, the spins of the Ising model are implemented using by approximating a probabilistic magnetic tunnel junction model \cite{pbit}.
Behaviors of the model can be approximated using integral stochastic computing, thus the probabilistic bit (p-bit) can be implemented by standard CMOS digital circuits \cite{CIL}.
Binary stochastic computing is a probabilistic computing system that represents a value using the probability of a `1' in a stochastic bit-stream \cite{sto_ori1,sto_ori2}.
Let us denote by $P_x$ the existence probability of `1' in the stochastic bit-stream $x \in \{0, 1\}$.
The value, $X \in [-1 : +1]$, which is represented by the bit-stream, is given by $2 \cdot P_x - 1$.
Integral stochastic computing extends the range of represented values to integers using multiple stochastic bit-streams \cite{sto_dnn}.
Stochastic computing has been applied to several applications such as deep neural networks \cite{sto_dnn} and image processing \cite{sto_img}.
\begin{figure}
    \centering
    \includegraphics[width=0.6\linewidth]{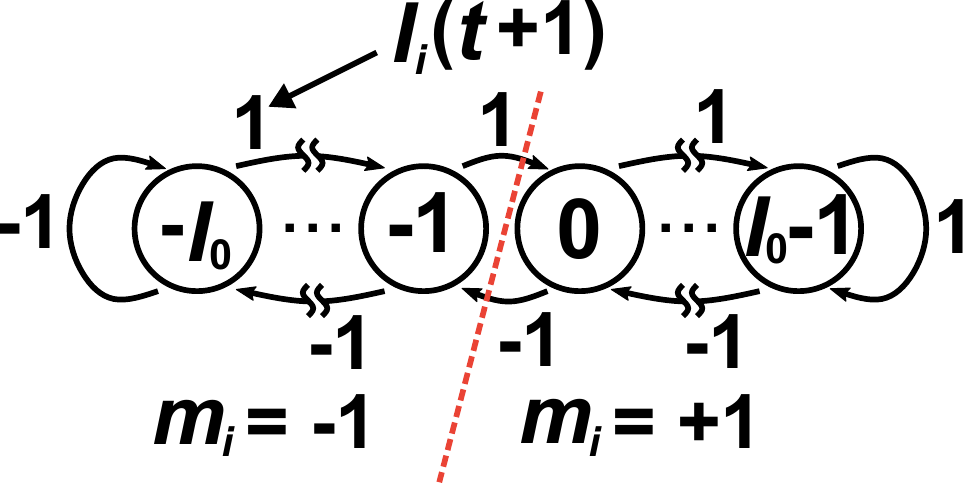}
    \caption{A finite state machine (FSM) for $\mathrm{Itanh}$ function.}
    \label{fig:f2}
\end{figure}
The behavior of a p-bit approximated by integral stochastic computing is given by:
\begin{subequations}
    \begin{equation}
        I_i (t + 1) = h_i + \sum_j J_{ij} m_j(t) + n_{rnd} \cdot r_i(t) + \mathrm{Itanh}_i (t) ,
        \label{eq:Ii}
    \end{equation}
    \begin{equation}
        \mathrm{Itanh}_{i}( t+1) =
            \begin{cases}
                I_{0}(t) - 1, & \mathrm{if} \ I_{i}( t+1) \geq I_{0}(t) ,\\
                -I_{0}(t), & \mathrm{else\ if } \ I_{i}( t+1) < -I_{0}(t) ,\\
                I_{i}(t+1), & \mathrm{otherwise},
            \end{cases}
        \label{eq:itanh}
    \end{equation}
    \begin{equation}
        \begin{aligned}
            m_i (t+1) & = \mathrm{sgn}(\mathrm{Itanh}_{i}(t+1)) \\
            & =
            \begin{cases}
                +1, & \text{ if } \: \mathrm{Itanh}_{i} (t+1) \ge 0,\\
                -1, & \text{ otherwise,}
            \end{cases} \\
        \end{aligned}
        \label{eq:mi}
    \end{equation}
    \label{eq:spin}
\end{subequations}
where $t$ is the number of cycles, $n_{rnd}$ is the magnitude of the noise signals, $r_i(t) \in \{-1, +1\}$ denotes the random noise signals, $\mathrm{sgn}$ is the sign function, and $I_0$ is the pseudoinverse temperature.
In \cref{eq:itanh} and (\ref{eq:mi}), $\mathrm{Itanh}$ is an approximated $\tanh$ function obtained using stochastic computing.
In stochastic computing, $\tanh$ function is calculated by a finite state machine (FSM), as shown in \cref{fig:f2}.
The conversion of binary numbers to the stochastic bit stream is not required, although the SSA algorithm is based on stochastic computing.
There are no input binary numbers for stochastic computing, and the output of spin is already `0' or `1' like the bit in the stochastic bit stream.
\begin{figure}
    \centering
    \includegraphics[width=0.9\linewidth]{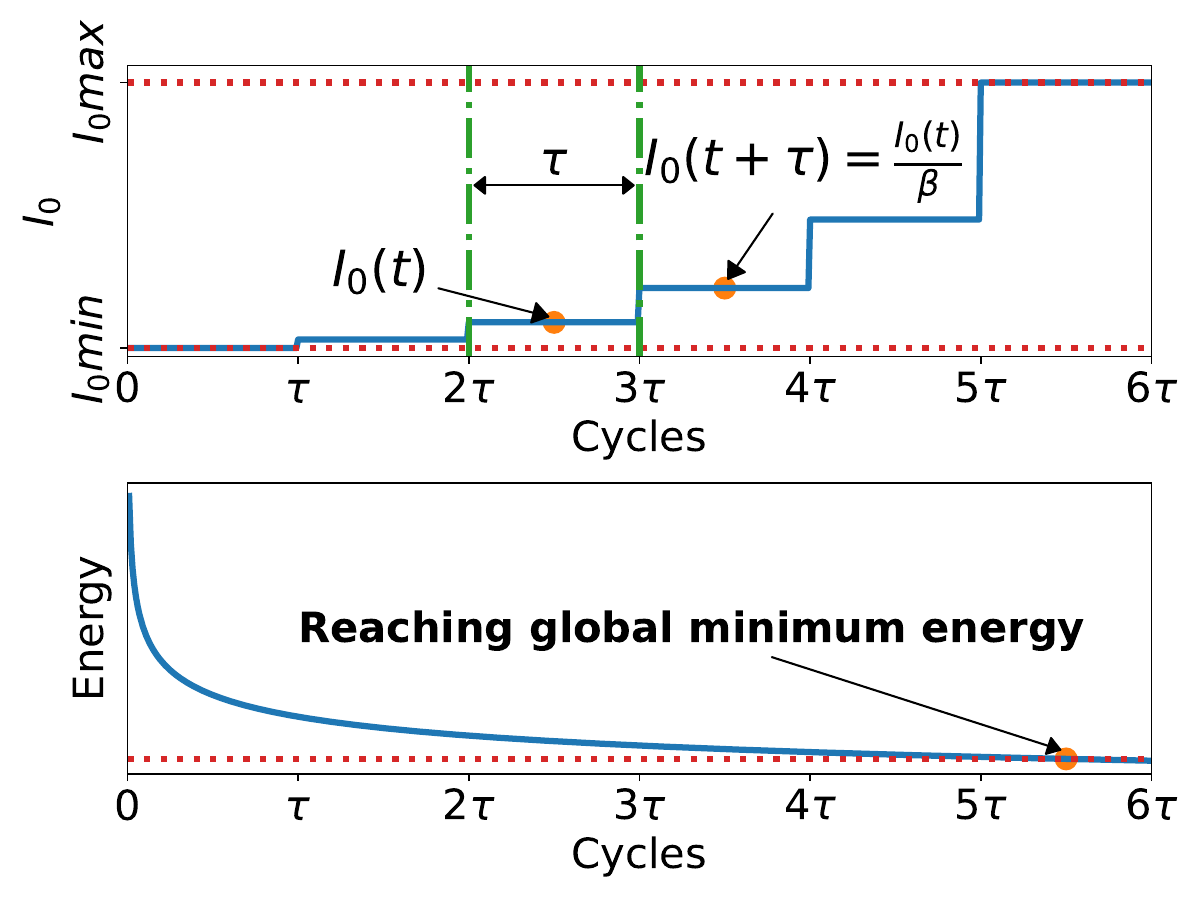}
    \caption{The pseudoinverse temperature $I_0$ of the proposed SSA algorithm, and the landscape of the Ising energy.}
    \label{fig:f3}
\end{figure}
In SSA, the spin fluctuation is controlled using the pseudoinverse temperature and several hyperparameters.
\cref{fig:f3} shows a landscape of the pseudoinverse temperature and the Ising energy versus the number of clock cycles.
The initial value of the temperature is defined as $I_{0min}$, and the temperature gradually increases to $I_{0max}$.
The pseudoinverse temperature is determined according to:
\begin{equation}
    I_0 (t + \tau) = \frac{I_0 (t)}{\beta}.
    \label{eq:tempr}
\end{equation}
Here, $\tau$ is the number in which $I_0$ has a stable value, and $\beta$ is the increase ratio of $I_0$.
When $I_0$ is small, the spin changes its state with high probability, so that the proposed hardware can traverse a wide search space.
In contrast, the spin becomes stable when $I_0$ is large; thus the Ising energy converges to the global minimum.
The SSA and SA algorithms are stochastic algorithms, therefore they do not obtain the optimal solution to the problem every time.
The average Ising energy of multiple trials, rather than the result of a single event, is compared to evaluate SSA and SA.
\subsection{MAX-CUT problem}
\label{sec:maxcut}
\begin{figure}[h]
    \centering
    \subfloat[]{
        \includegraphics[width=0.3\linewidth]{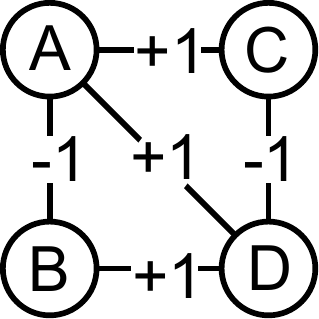}
        \label{fig:f5a}
    }
    \subfloat[]{
        \includegraphics[width=0.25\linewidth]{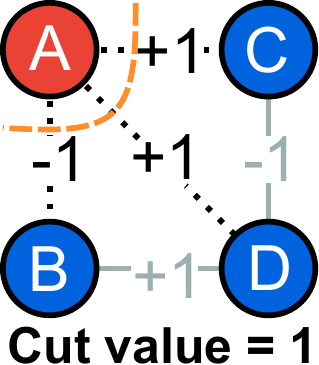}
        \label{fig:f5b}
    }
    \subfloat[]{
        \includegraphics[width=0.25\linewidth]{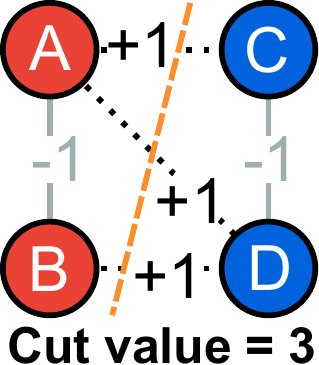}
        \label{fig:f5c}
    }
    \caption{The MAX-CUT problem. (a) An example of a MAX-CUT problem. (b) Incorrect solution. (c) Correct solution.}
    \label{fig:f5}
\end{figure}
The maximum cut (MAX-CUT) problem is one of the typical combinatorial optimization problems, and it is used for evaluation purposes in several studies \cite{tnnls, statica_512, 20k_spin}. 
A MAX-CUT problem is a graph partitioning problem that maximizes the sum of the edge's weights between two partitioned graphs \cite{maxcut, maxcut1}.
Let us denote by $G=(V,E)$ a graph in which $V=\{1,2,...,N\}$, $E \subset \{(i, j) : 1 \le i < j \le N\}$, and $N=|V|$.
The MAX-CUT problem finds the bi-partition $(V_1, V_2)$ of $V$ such that the sum of the edge weights between $V_1$ and $V_2$ is maximized.
\cref{fig:f5} (a) shows a 4-vertex MAX-CUT problem where the edge weights are only `+1' or `-1'.
When the graph is divided into $V_1 = \{A\}$ and $V_2 = \{B, C, D\}$, as in \cref{fig:f5} (b), the sum of the cut edge weights (the cut value) is 1.
On the other hand, the cut value of \cref{fig:f5} (c) is 3, so the optimal solution of this problem includes $V_1 = \{A, B\}$ and $V_2 = \{C, D\}$.
\begin{table}
    \caption{Summary of the benchmark problems.}
    \label{tb:t1}
    \begin{center}
        \begin{tabular}{|c|c|c|c|c|}
            \hline
            Problem & \# vertices & \# edges & Weights & Best known \\
            \hhline{|=|=|=|=|=|}
            G11 & 800 & 1,600 & \{-1, 1\} & 564 \\
            \hline
            G12 & 800 & 1,600 & \{-1, 1\} & 556 \\
            \hline
            G13 & 800 & 1,600 & \{-1, 1\} & 582 \\
            \hline
            King1 & 800 & 3,200 & \{-1, 1\} & N/A \\
            \hline
        \end{tabular}
    \end{center}
\end{table}
There is a benchmark dataset for the MAX-CUT problem, as same as other optimization problems.
The dataset of the MAX-CUT problem is `G-set' \cite{G-set}, which contains 71 MAX-CUT problems.
The structure of the problems is categorized into three categories: random, planar, and toroidal graph, however, the weights of the problems are only `+1' or `-1'.
In this paper, G11, G12, and G13, which are problem instances from the G-set \cite{G-set}, are used for the evaluations.
In addition, the custom MAX-CUT problem, which has more edges than G11, G12, and G13, is used.
The custom problem is 'King1', and the vertices of this problem are connected to 8 neighborhood vertices.
\cref{tb:t1} summarizes the benchmark problems that are used for the evaluations.
Each vertex of G11, G12, and G13 is connected to 4 neighborhood vertices; thus, the number of edges is 1,600.
King1 is the custom-made MAX-CUT problem for evaluating whether the proposed HA-SSA algorithm can be applied to more complex MAX-CUT problems.
The topology of King1 is King`s graph \cite{kings}, so that, each vertex of King1 is connected to 8 neighborhood vertices.
It has 3,200 weights, which are distributed according to the uniform distribution.
\section{HA-SSA algorithm}
\label{sec:hassa}
\subsection{Pseudoinverse temperature control}
\label{sec:tem}
The pseudoinverse temperature is controlled by five hyperparameters in SSA.
$I_{0min}$ is the initial temperature value, and $I_{0max}$ is the maximum value.
$\tau$ is the number of cycles that maintains the same temperature value.
$\beta$ is the temperature increase ratio, and $n_{rnd}$ denotes the magnitudes of the random noise signals in \cref{eq:itanh}.
In \cite{tnnls}, the optimal combinations of the hyperparameters except $\beta$ were found for both real-numbered and integer representations.
SSA can obtain high-quality solutions with high probability even when using only the integer representations of $I_{0min}$, $I_{0max}$, and $n_{rnd}$.
However, $\beta$ is represented by a real number, and the pseudoinverse temperature is determined by \cref{eq:tempr}.
If \cref{eq:tempr} is directly implemented in the hardware, a floating point divider is required.
The floating-point representation requires considerable hardware costs, and it represents a hardware bottleneck.
For hardware-friendly temperature control, $\beta$ is represented by an integer rather than a floating-point in the proposed HA-SSA algorithm.
Additionally, \cref{eq:tempr} is modified to:
\begin{equation}
    I_0 (t + \tau) = 2^\beta \cdot I_0(t).
    \label{eq:hd_tempr}
\end{equation}
With \cref{eq:hd_tempr}, the multiplication of $2^\beta$ and $I_0(t)$ is performed by shifting $I_0(t)$; thus, the hardware cost can be significantly reduced.
Additionally, all hyperparameters are represented by integers only.
In \cite{tnnls}, the duration of the SSA process is controlled by the number of cycles.
One iteration is regarded as a process while the pseudoinverse temperature increases from $I_{0min}$ to $I_{0max}$.
This method can cause the entire computation procedure to stop before the temperature reaches its maximum value in the last iteration.
For instance, assume that $I_{0min}$ is 1, $I_{0max}$ is 32, $\beta$ is 1 and $\tau$ is 100.
In this case, one iteration contains 6 steps and its procedure includes 600 cycles.
If the entire computation procedure includes 10,000 cycles, the last iteration is stopped during its process.
To prevent this problem, HA-SSA controls the duration of the computation process via the number of iterations.
The temperature $I_0(t)$ for SSA is determined by \cref{eq:tempr}.
When $\beta$ in \cref{eq:tempr} is 0.5 and $\beta$ in \cref{eq:hd_tempr} is 1, the temperature control of HA-SSA is the same as that of SSA.
In addition, the calculations of the spin states in HA-SSA are the same as SSA, thus, the annealing results of the HA-SSA and SSA algorithms are the same when the hyperparameters of HA-SSA and SSA are equivalent.
\subsection{Memory efficiency improvement}
\label{sec:mem}
The SSA algorithm only calculates the probability of the update for each spin rather than calculating the Ising energy defined by \cref{eq:ham}, because it is based on a Boltzmann machine \cite{BM}.
This is considered one reason that the SSA algorithm converges faster than the conventional SA algorithm.
However, since SSA does not calculate the Ising energy, it is uncertain when the energy converges to the global minimum.
The spin states should be stored for determining when the Ising energy is converged to the global minimum without a complex energy-computation circuit.
Furthermore, the spin states have to be stored, because the solution of the combinatorial optimization problem is the spin states, not the Ising energy.
To find the solution to the problem, the annealing hardware requires memory for storing the spin states, even if it calculates the energy inside it.
The memory usage of SSA for storing the spin states in one iteration, $M$, can be represented by:
\begin{equation}
    M = N \cdot (\log_\beta \frac{I_{0min}}{I_{0max}} + 1) \cdot \tau \text{  bits},
    \label{eq:cv_mem}
\end{equation}
where $N$ is the number of spins.
$M$ is determined by the hyperparameters and the number of spins.
For example, let us denote by $I_{0min}=1$, $I_{0max}=16$, $\beta=1$, and $\tau=100$.
One iteration has 5 steps, thus the number of cycles is 500 cycles in one iteration.
In this example, one iteration requires $500N$ bits of memory to store the entire spin states during one iteration.
The proposed HA-SSA hardware reduces the memory requirements by storing only a subset of spin states according to the pseudoinverse temperature, rather than storing all spin states.
In the SA algorithm, the spin state converges to the near-optimal state with a high probability when the temperature is low.
Therefore, if $I_0max$ is sufficiently large, the lower Ising energy appears with a high probability when the pseudoinverse temperature is high.
By storing the result obtained when the pseudoinverse temperature is the maximum ($I_0(t) = I_{0max}$), HA-SSA realizes memory-efficient annealing hardware.
Because HA-SSA reduces the memory usage by determining the stored spin state, it is not just a memory-efficient hardware design.
The memory usage of HA-SSA in one iteration, $M'$, is defined by:
\begin{equation}
    M' = N \cdot \tau \text{   bits}.
    \label{eq:pr_mem}
\end{equation}
$M'$ is only determined by $\tau$; thus HA-SSA can achieve $(\log_\beta (I_{0min} / I_{0max}) + 1)$ times more memory efficiency.
\section{Experimental environment}
\label{sec:setup}
\subsection{Simulation setup}
\label{sec:sim}
\begin{table}
    \caption{The combination of the hyperparameter used for HA-SSA.}
    \label{tb:t2}
    \begin{center}
        \begin{tabular}{|c|c|c|c|c|c|c|}
            \hline
            $trial$ & $m_{shot}$ & $n_{rnd}$ & $I_{0min}$ & $I_{0max}$ & $\tau$ & $\beta$ \\
            \hhline{|=|=|=|=|=|=|=|}
            100 & 150 & 2 & 1 & 32 & 100 & 1 \\
            \hline
        \end{tabular}
    \end{center}
\end{table}
Before implementing the HA-SSA hardware, the HA-SSA algorithm is evaluated by simulations and compared to the conventional SSA and SA algorithms.
\cref{tb:t2} shows the hyperparameter combination used for HA-SSA.
The temperature of HA-SSA increases according to \cref{eq:hd_tempr}, thus the equivalent $\beta$ for conventional SSA is 0.5.
$m_{shot}$ is the number of iterations in each trial.
Therefore, the entire process of simulation for one trial is 90,000 cycles, because one iteration is 600 cycles with these parameters.
Similar to the proposed method, the conventional SSA method simulates 90,000 cycles.
Additionally, the SA algorithm is applied to the same problems for the comparisons.
The temperature of SA gradually decreases from 10 to $10^{-7}$ during the 90,000 cycles.
The SSA simulations and the SA method are conducted using Python on a system with a 16-core Intel i9-12900KF CPU at 3.2 GHz and 64-GB of memory.
\subsection{Design of the HA-SSA hardware}
\label{sec:hard}
\begin{figure}[h]
    \begin{center}
        \includegraphics[width=\linewidth]{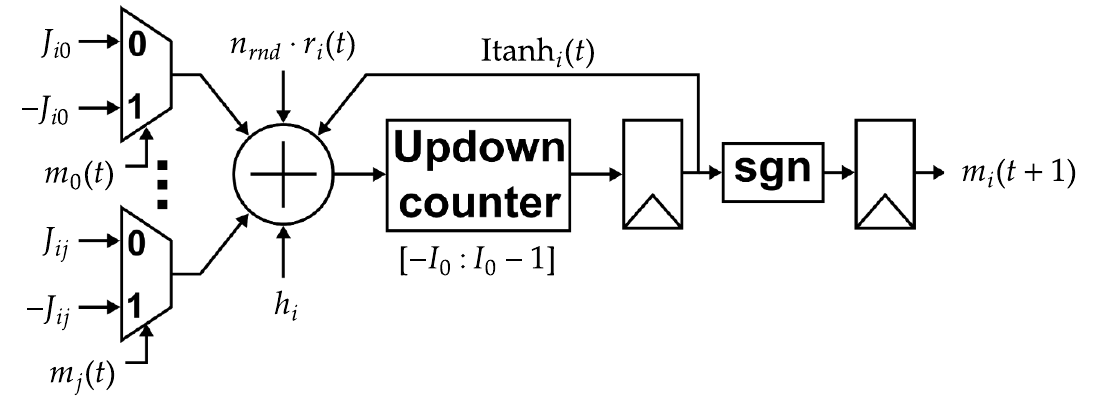}
        \caption{Block diagram of the spin-gate circuit.}
        \label{fig:f6}
    \end{center}
\end{figure}
Based on the proposed HA-SSA algorithm, the annealing hardware is implemented on an FPGA.
In HA-SSA, the spin of the Ising model is implemented using the p-bit based on stochastic computing.
The calculations of the p-bit given by \cref{eq:Ii}, (\ref{eq:itanh}) and (\ref{eq:mi}) are implemented using the spin-gate circuit shown in \cref{fig:f6}.
Because the spin-gate is based on stochastic computing, the multiplication of $J_{ij}$ and $m_j$ is calculated using a two-input multiplexer, and the addition operation is performed using a binary adder.
As mentioned before, \cref{eq:itanh} is realized using the FSM shown in \cref{fig:f2}, and the FSM is easily implemented using a saturated up-down counter.
The number of FSM states is $2 \cdot I_0$, and $m_i$ is `+1' when $\mathrm{Itanh}(I_i) \ge 0$, otherwise $m_i$ is `-1'.
However, the output of the spin-gate is a stochastic bit stream, so $m_i = -1$ is represented as the logical value ``0'', and $m_i = +1$ is represented as the logical value ``1''.
Similar to $m_i$, $r_i(t)$ in \cref{eq:Ii} is represented as ``0'' when $r_i(t)$ is `-1' and ``1'' when $r_i(t)$ is `+1'.
The spin-gate circuit can also be used for CMOS invertible logic (CIL) which provides a capability of probabilistic bidirectional operation between inputs and outputs of a function \cite{CIL}.
In the CIL, the spin-gates for the inputs or the outputs of the function have to be fixed its state to determine the unfixed spin state correctly for the desired function.
Therefore, the spin-gate in \cite{CIL} needs a capability that determines whether the spin-gate is a fixed or unfixed mode.
For implementing the CIL circuit, the desired function should be converted to the Ising model, however, the Ising model in the CIL has a sparse topology.
In \cite{sparseCIL}, the spin-gate has scalability by using the sparse topology of the Ising model for the CIL.
In contrast, the Ising model converted from the combinatorial optimization problems can be fully connected as well as sparse.
The spin-gate in HA-SSA should be able to support the fully-connected Ising model.
\begin{figure}
    \begin{center}
        \includegraphics[width=0.95\linewidth]{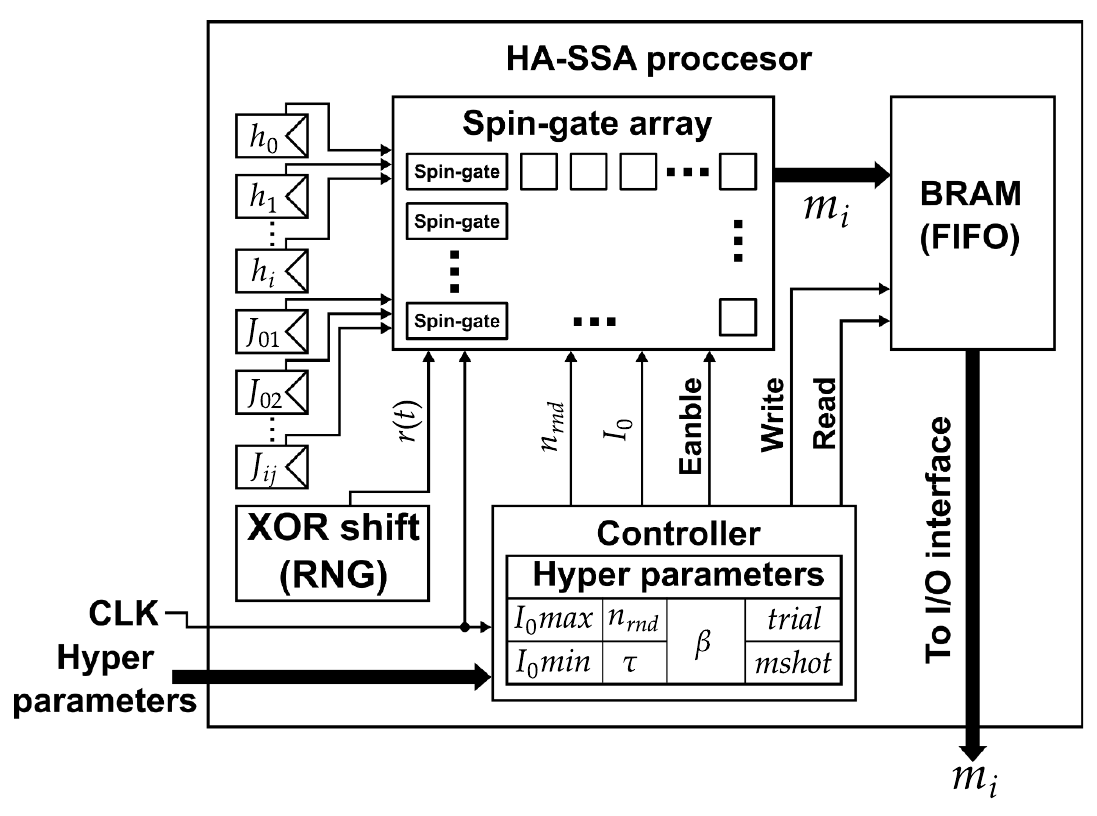}
        \caption{Block diagram of the HA-SSA hardware.}
        \label{fig:f7}
    \end{center}
\end{figure}
\cref{fig:f7} shows the architecture of the proposed HA-SSA hardware.
It contains a spin-gate array, a random number generator (XOR-shift) \cite{xorshift}, a controller, registers for the biases and the weights, and a first in, first out (FIFO) memory using block random access memory (BRAM).
The spin-gate array contains the spin-gates and the spin-gates are connected to each other.
The number of the spin-gates and their connections are determined by the Ising model.
The spin-gate array receives $h_i$ and $J_{ij}$ from the registers located at the top module.
In each clock cycle during the annealing process, the spin-gate array produces $m_i(t+1)$ signals as result of the annealing process.
The results produced by the spin-gate array are stored in the FIFO BRAM.
The width of the BRAM depends on the number of spin-gates.
The random bit signals, $r(t)$ in \cref{eq:Ii}, are generated using the XOR-shift random number generator (RNG).
The bit width of $r(t)$ is the same as the number of the spin-gates, because each spin-gate in the array needs a random noise bit for its stochastic operations.
The controller controls the spin-gate array, the random number generator, and the BRAM.
It controls the spin-gate array using several control signals such as an enable signal.
It generates the pseudoinverse temperature, $I_0$, according to \cref{eq:hd_tempr} by using the received hyperparameter ($I_{0min}$, $I_{0max}$, $\tau$, and $\beta$).
Additionally, the BRAM is controlled by the controller.
In HA-SSA, the results from the spin-gate array are stored when $I_0$ is the maximum; thus, the controller generates a write signal according to the value of $I_0$.
\subsection{FPGA Implementation}
\label{sec:fpga}
\begin{table}
    \caption{Summary of the HA-SSA hardware used for the G11 and King1 problems.}
    \label{tb:t3}
    \begin{center}
        \begin{tabular}{|c|c|c|c|}
            \hline
            \multicolumn{2}{|c|}{} & G11 & King1 \\
            \hhline{|==|=|=|}
            \multicolumn{2}{|c|}{Clock frequency} & \multicolumn{2}{c|}{100 MHz} \\
            \hline
            \multirow{3}{2cm}{FPGA resource usage} & LUT & 105,294 (51.67\%) & 170,035 (83.43\%) \\
            \cline{2-4}
            & FF & 13,692 (3.36\%) & 14,738 (3.62\%) \\
            \cline{2-4}
            & BRAM & \multicolumn{2}{c|}{356 (80.00\%)} \\
            \hline
            \multicolumn{2}{|c|}{Power dissipation} & 2.138 W & 5.532 W \\
            \hline
            \multicolumn{2}{|c|}{\# of spin-gates} & \multicolumn{2}{c|}{800} \\
            \hline
            \multicolumn{2}{|c|}{\# of weight registers} & 1,600 & 3,200 \\
            \hline
        \end{tabular}
    \end{center}
\end{table}
The HA-SSA hardware used for the G11 and King1 MAX-CUT problems is implemented on a Digilent Genesys 2 FPGA board which is powered by a Xilinx Kintex-7(XC7K325T-2FFG900C).
The FPGA resource usage and the power dissipation are summarized in \cref{tb:t3}.
Both problems have 800 vertices; thus, the proposed hardware contains 800 spin-gates.
Each vertex in the G11 problem is connected to 4 vertices so that the spin-gate shown in \cref{fig:f6} contains 4 multiplexers.
The number of weight registers is 1,600.
In contrast, the King1 problem has 8 connections at each vertex; thus, the number of weights is 3,200.
The look-up table (LUT) utilization rates for G11 and King1 are 51.67\% and 83.43\%, respectively.
The width of the FIFO BRAM, which stores the results, is 800 bits because the number of spin-gates is 800.
When the width is 800 bits, the maximum possible depth of the BRAM is 16,384.
The proposed hardware receives the hyperparameters from the PC, and it returns its annealing results to the PC.
The hardware communicates with the PC via a universal asynchronous receiver/transmitter (UART) interface.
The PC employed for communication is the same one that is used in the algorithm simulations.
\section{Evaluation}
\label{sec:eval}
\subsection{Algorithm evaluation}
\label{sec:alg_eval}
\begin{figure}
    \begin{center}
        \includegraphics[width=0.9\linewidth]{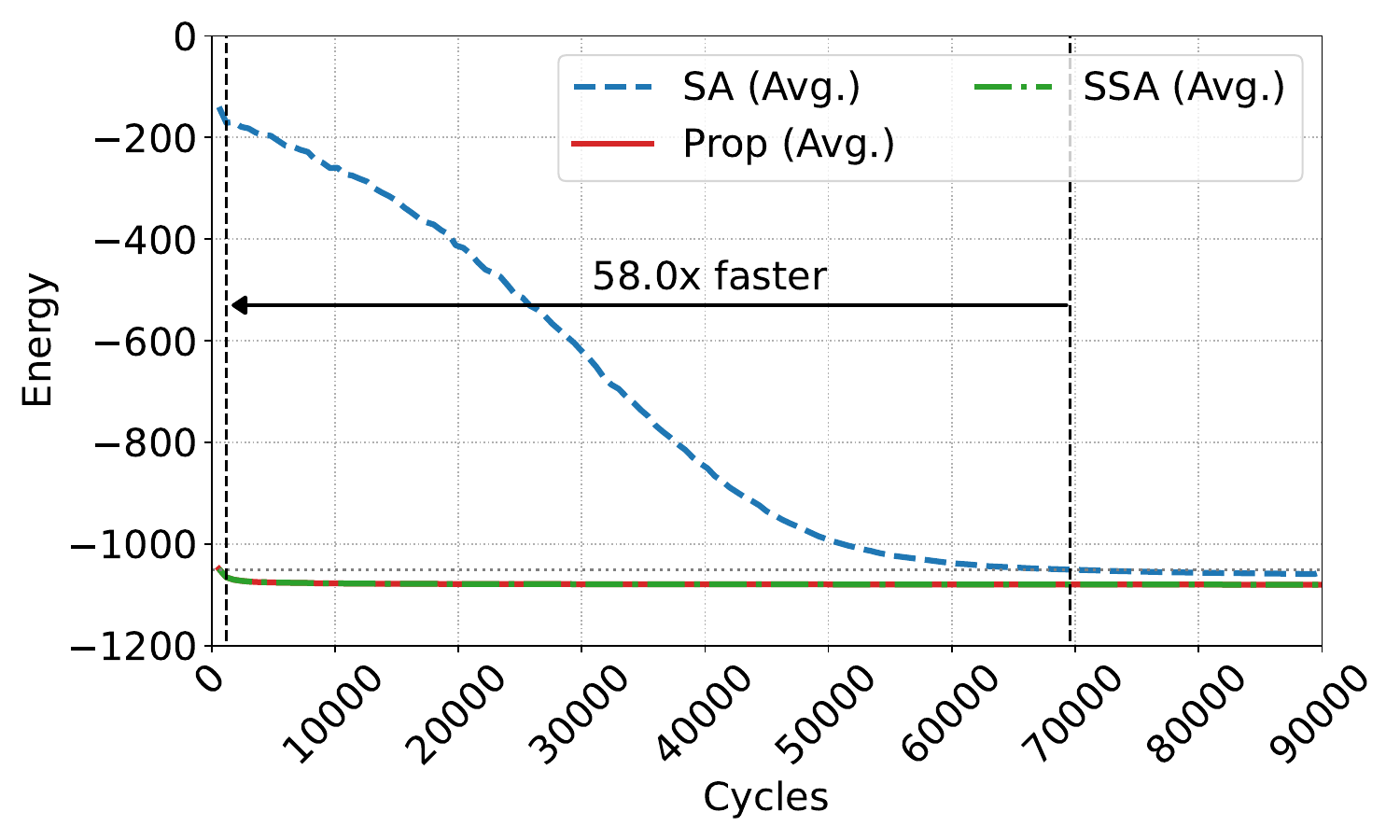}
        \caption{Average Ising energy values for G11 versus the number of cycles.}
        \label{fig:f8}
    \end{center}
\end{figure}
The MAX-CUT problems (G11, G12, and G13) are solved using the proposed HA-SSA method, the conventional SSA method \cite{tnnls}, and the SA method \cite{ising_sa}.
\cref{fig:f8} shows the average Ising energy values versus the number of cycles for G11 during 100 trials.
Although the results of the proposed method are only stored when the temperature is the maximum value, it achieves the same results as those of the conventional SSA method.
The SA, SSA, and HA-SSA algorithm are the stochastic algorithms so that the convergence speed to the near-optimal solution of each method are compared.
The SA algorithm achieves -1050 of the Ising energy, which is 96\% of the best known energy for G11, within 69,600 cycles.
In contrast, the proposed method achieves the same Ising energy within 1,200 cycles which is 58-times faster than that of the SA algorithm.
In addition, the proposed method achieves its best average Ising energy (-1,079.5) within 75,200 cycles.
The best average Ising energy from SA is -1,058.6, and it is achieved at 89,400 cycles, thus the proposed method converges 1.19-times faster than SA.
\begin{figure*}
    \begin{center}
        \includegraphics[width=0.9\linewidth]{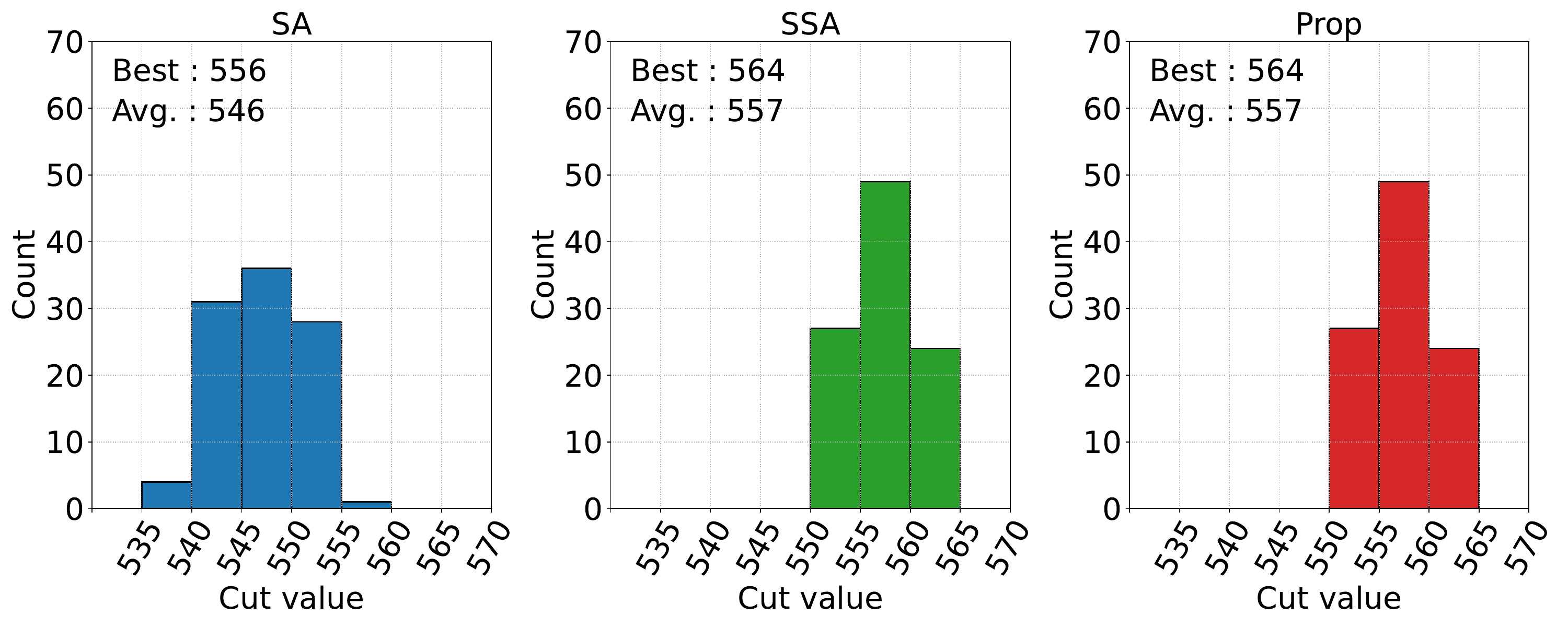}
        \caption{Histogram of the maximum cut values obtained during 100 trials of G11.}
        \label{fig:f9}
    \end{center}
\end{figure*}
\cref{fig:f9} shows the histograms of the maximum cut values produced by each method during 100 trials when solving G11.
The conventional SSA method obtains the best known solution of G11, and it also obtains a higher average value than that of SA.
The proposed method achieves the same best and average cut values for 100 trials; thus, the proposed HA-SSA algorithm is equivalent to the conventional SSA approach.
\begin{figure}[]
    \begin{center}
        \subfloat[]{
            \includegraphics[width=0.9\linewidth]{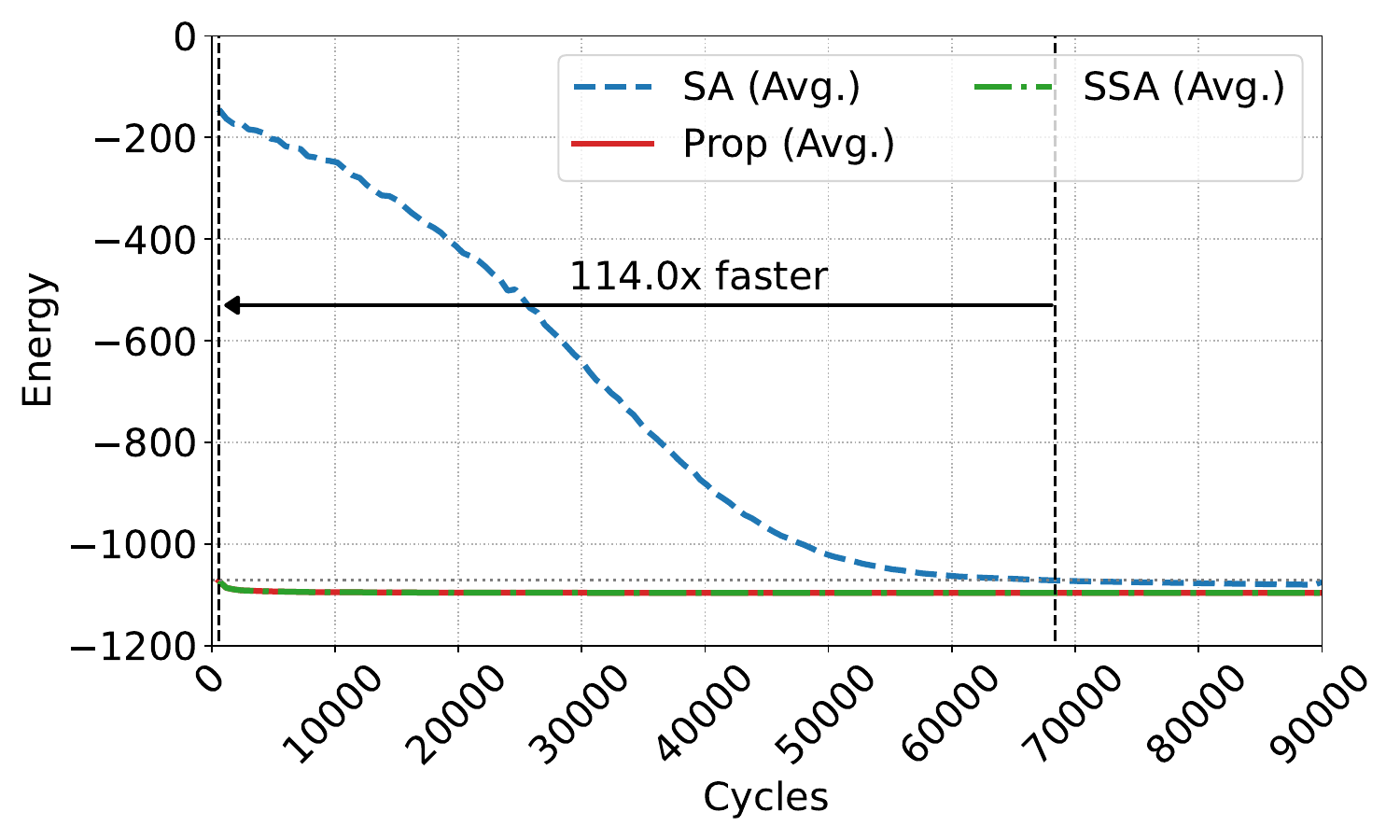}
            \label{fig:f10a}
        }
        \hfill
        \subfloat[]{
            \includegraphics[width=0.9\linewidth]{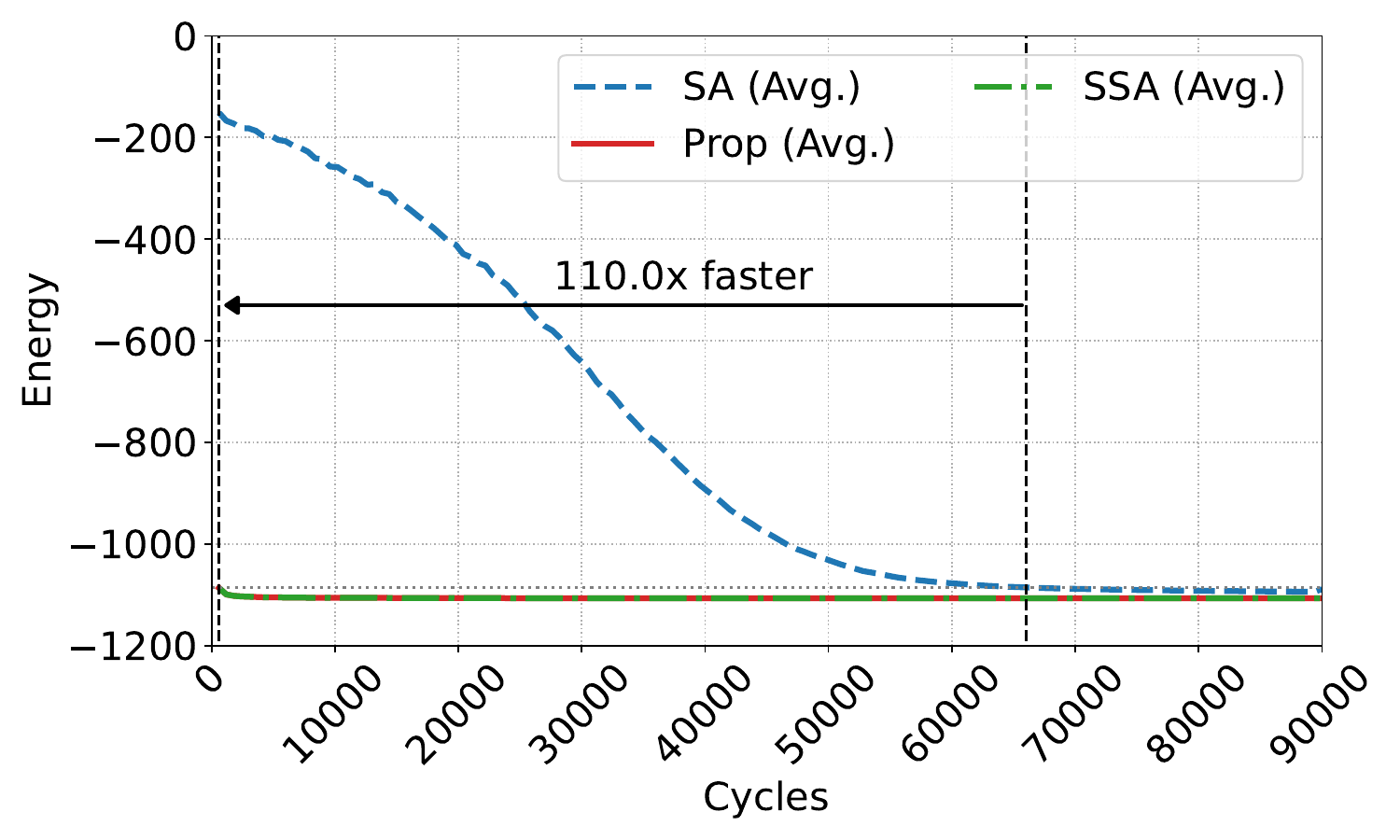}
            \label{fig:f10b}
        }
        \caption{Average Ising energy values versus the number of cycles. (a) G12. (b) G13.}
        \label{fig:f10}
    \end{center}
\end{figure}
\begin{figure*}[]
    \begin{center}
        \subfloat[]{
            \includegraphics[width=0.9\linewidth]{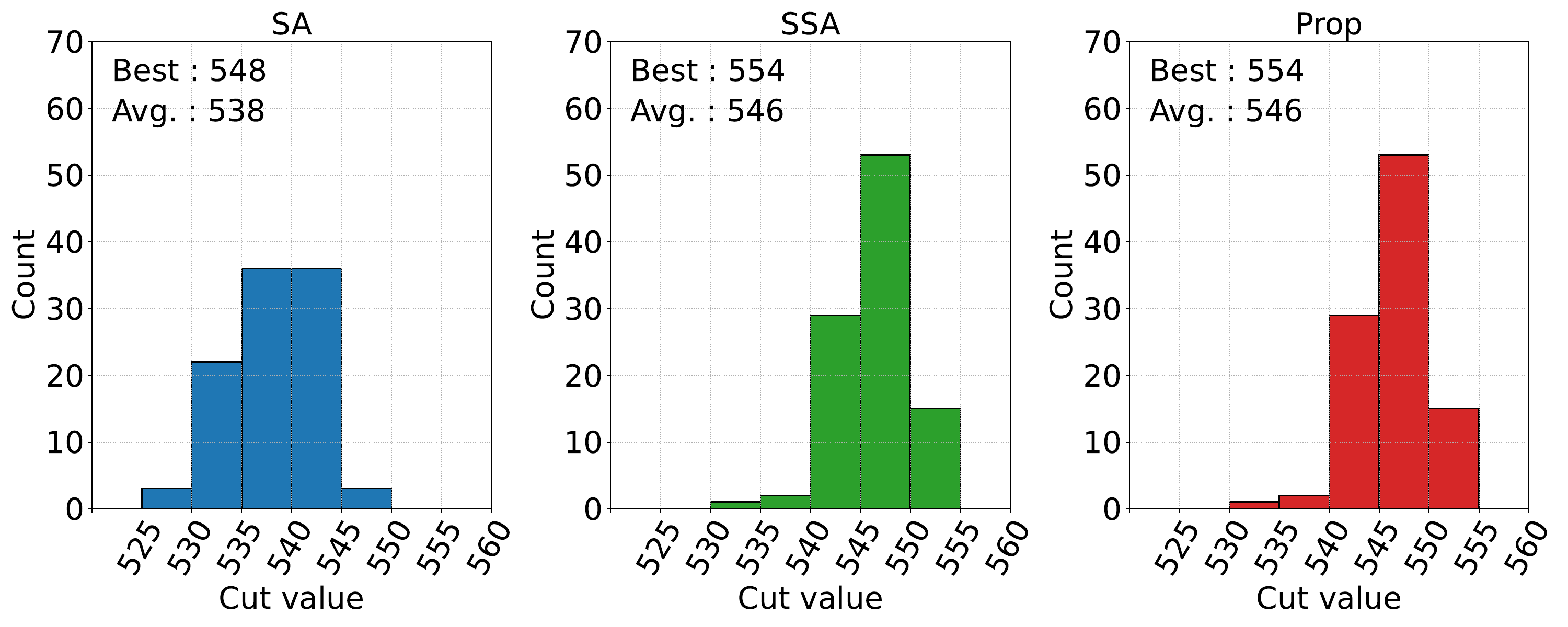}
            \label{fig:f11a}
        }
        \hfill
        \subfloat[]{
            \includegraphics[width=0.9\linewidth]{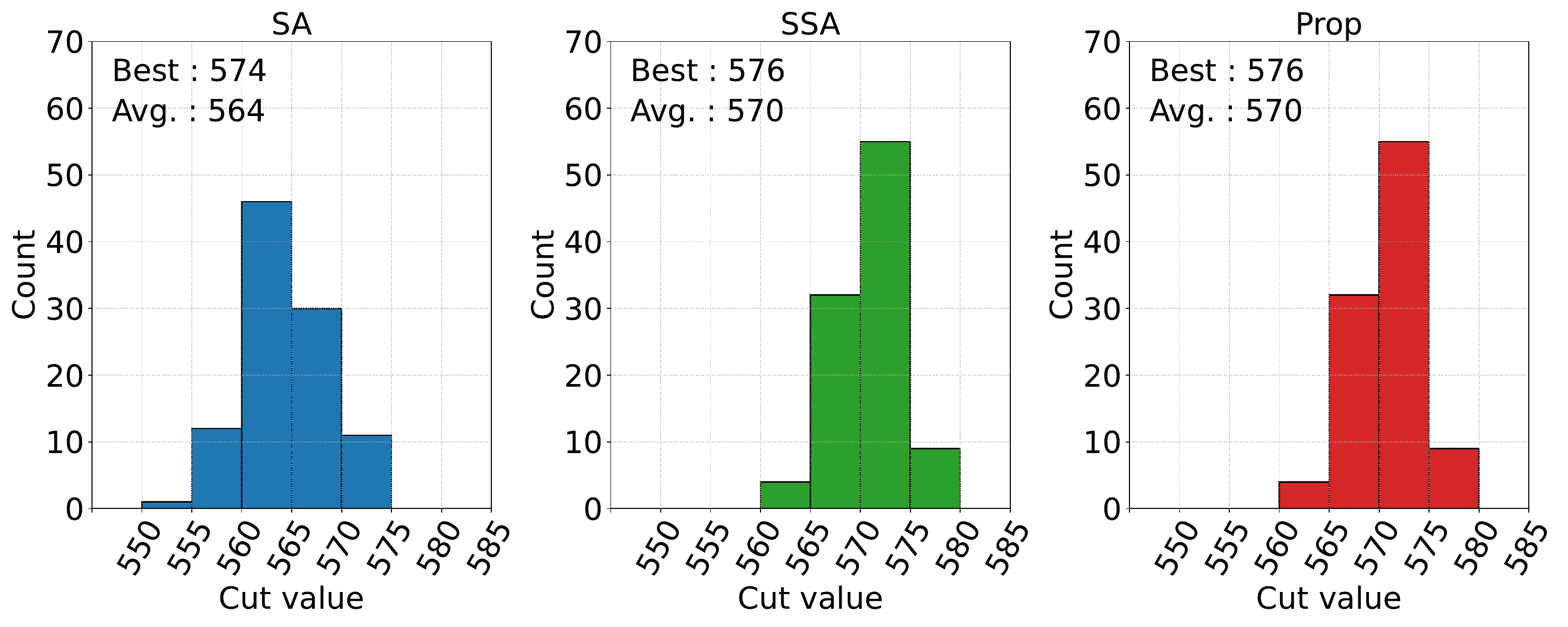}
            \label{fig:f11b}
        }
        \caption{Histograms of the maximum cut values obtained during 100 trials. (a) G12. (b) G13.}
        \label{fig:f11}
    \end{center}
\end{figure*}
\cref{fig:f10} (a) and (b) show the average energy values versus the number of cycles for G12 and G13, respectively.
As in G11, the proposed method produces the same results as the conventional method.
First, the SA algorithm achieved 96\% of the best known solution for G12 within 68,400 cycles, however, the proposed method achieves the same Ising energy within 600 cycles which is 114-times faster than that of SA.
The best average Ising energy from the proposed method is -1,095.9 which is 98.2\% of the best known solution, and it is achieved within 82,800 cycles.
In contrast, the SA algorithm achieves -1,079.7 of the Ising energy within 89,400 cycles.
Same to the result of G12, the proposed method achieves 96\% of the best known solution within 600 cycles that is 110-times faster than where the SA achieves the same Ising energy.
The proposed HA-SSA method achieves the best average Ising energy (-1,106.5) during 71,400 cycles, however, the conventional SA method achieves its best Ising energy (1,093.6) during 89,400
The hyperparameter of the annealing process are equal to those in \cref{tb:t2}, so the proposed method can achieve an equivalent solution to that of the conventional method with 6 times its memory efficiency.
\cref{fig:f11} (a) and (b) show histograms of the cut values obtained during 100 trials of G12 and G13.
The proposed method achieves higher best and average solutions than the SA algorithm in both problems.
However, the best solutions of the proposed method are not the best known solutions for both problems.
The hyperparameters which are used for solving G12 and G13 are equal to the combination in \cref{tb:t2}.
However, this combination is optimized for G11; thus, they may not be optimal for G12 and G13.
\begin{table}
    \caption{Memory usage and cut values comparison between HA-SSA and SSA for each problem.}
    \label{tb:t4}
    \begin{center}
        \begin{tabular}{|c|c|c|c|c|c|c|}
            \hline
            & \multicolumn{3}{c|}{SSA \cite{tnnls}} & \multicolumn{3}{c|}{HA-SSA} \\
            \hline
            Problem & Best & Avg. & $M$ & Best & Avg. & $M'$ \\
            \hhline{|=|=|=|=|=|=|=|}
            G11 & 564 & 557 & \multirow{3}*{0.48 M} & 564 & 557 & \multirow{3}*{0.08 M} \\
            \hhline{|-|--~|--~|}
            G12 & 554 & 546 & & 554 & 546 & \\
            \hhline{|-|--~|--~|}
            G13 & 576 & 570 & & 576 & 570 & \\
            \hline
        \end{tabular}
    \end{center}
\end{table}
\cref{tb:t4} summarizes the best and average cut value and the memory usage of SSA and HA-SSA.
The memory usage of SSA calculated by \cref{eq:cv_mem} is 0.48 Mega-bits per iteration; thus, one trial (150 iterations) consumes 72 M-bits.
In contrast, the memory usage of HA-SSA only is changed by $\tau$ value according to \cref{eq:pr_mem}; the proposed hardware consumes 12 M-bits per trial, making it 6 times more memory-efficient than SSA.
HA-SSA achieves exactly the same result as SSA with reduced memory usage.
\subsection{HA-SSA hardware evaluation}
\label{sec:hard_eval}
\begin{figure}[]
    \begin{center}
        \subfloat[]{
            \includegraphics[width=0.9\linewidth]{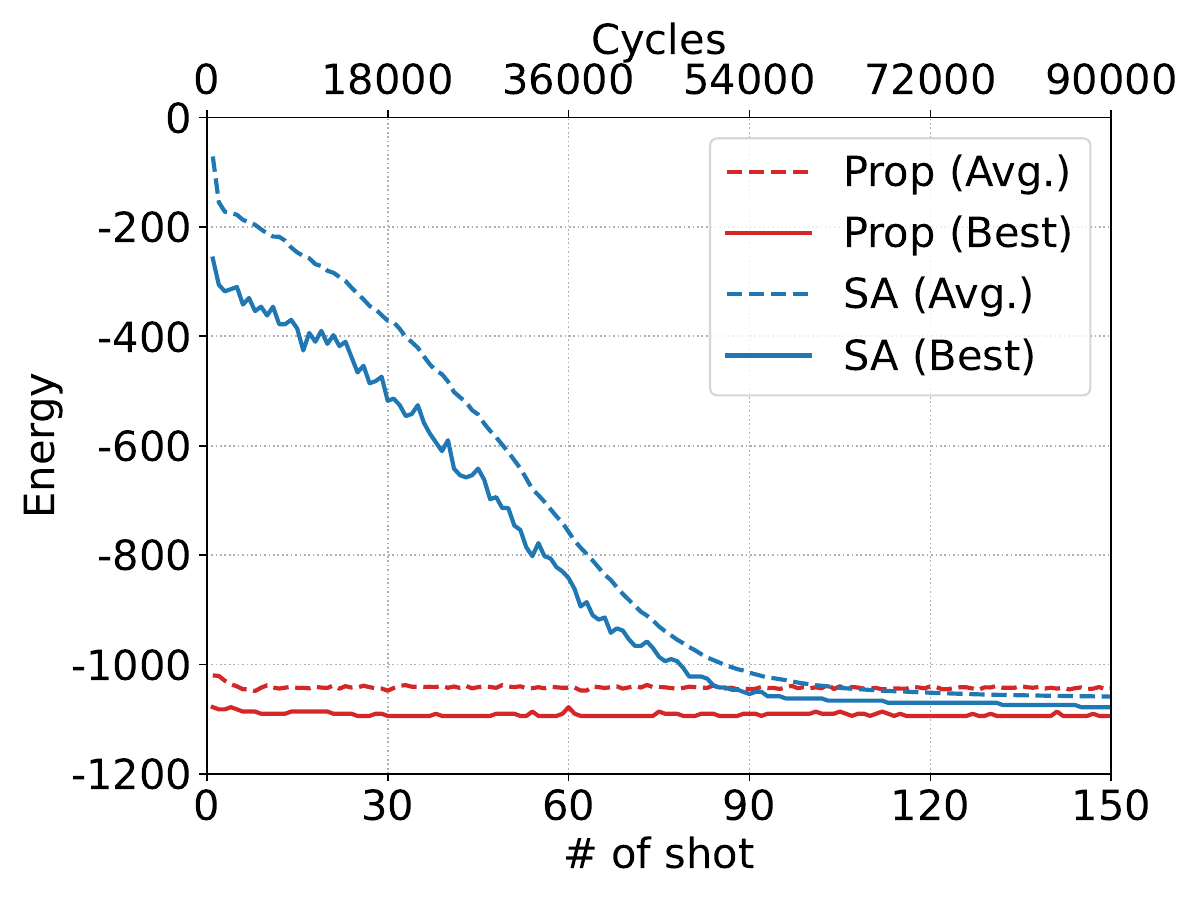}
        }
        \hfill
        \subfloat[]{
            \includegraphics[width=0.9\linewidth]{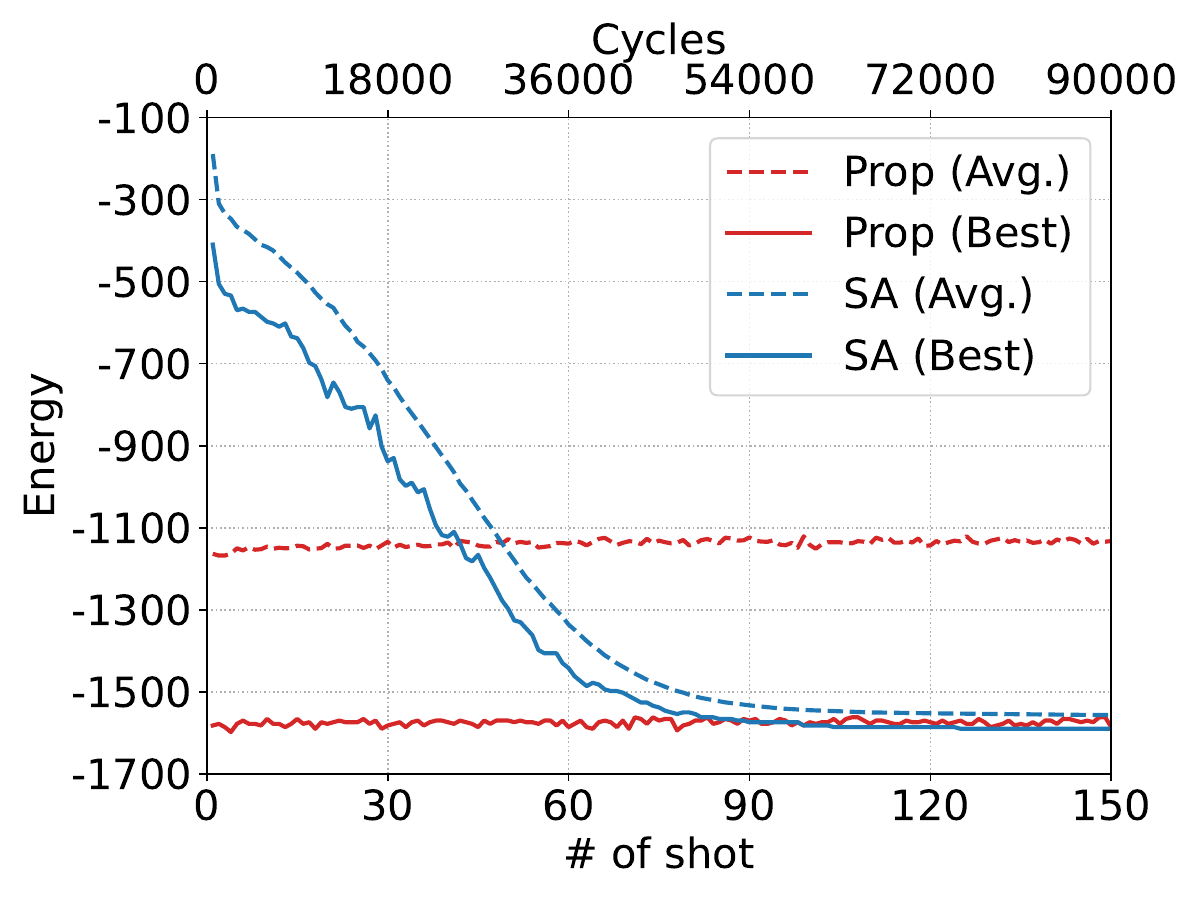}
        }
        \caption{Best and average Ising energy values at each iteration. (a) G11. (b) King1.}
        \label{fig:f12}
    \end{center}
\end{figure}
The proposed annealing hardware based on HA-SSA solves G11 and King1 for evaluation purposes, and this approach is compared to the SA algorithm (performed by the CPU).
As mentioned before, the calculation of the SSA algorithm is equivalent to that of HA-SSA.
The main difference between these two algorithms is memory usage during the annealing process.
It means that there is no difference between the results of HA-SSA hardware and SSA hardware when they use the same hyperparameters.
Therefore, the Ising energy obtained by the HA-SSA hardware is compared with the results of the SA algorithm, instead of SSA.
\cref{fig:f12} (a) and (b) show the best and average Ising energy values obtained by HA-SSA and conventional SA at each iteration for G11 and King1, respectively.
The HA-SSA hardware almost converges the energy from the first iteration for both problems.
It can achieve the best known solution for G11, although the hardware only stores the result obtained when $I_0$ is the maximum value.
As in G11, the proposed hardware converges faster than SA, and it achieves a higher-quality solution than SA.
\begin{table}[]
    \caption{Comparisons between the proposed hardware and the SA algorithm.}
    \label{tb:t5}
    \begin{center}
        \begin{tabular}{|c|c|c|c|c|}
            \hline
            Problem & Method & Annealing time & Best & Avg. \\
            \hhline{|=|=|=|=|=|}
            \multirow{2}*{G11} & SA & 342.29 s & 556 & 547 \\
            \cline{2-5}
            & Prop & 0.9 ms & 564 & 558 \\
            \hhline{|=|=|=|=|=|}
            \multirow{2}*{King1} & SA & 219.14 s & 812 & 795 \\
            \cline{2-5}
            & Prop & 0.9 ms & 816 & 806 \\
            \hline
        \end{tabular}
    \end{center}
\end{table}
\cref{tb:t5} shows comparisons between the proposed hardware and the SA algorithm.
The SA algorithm achieves 547 as the average cut value for G11, which is 96.99\% of the best known solution, during 100 trials.
The computation time of SA is the average calculated over 100 trials, and each trial is operated for 90,000 cycles.
In contrast, the proposed hardware achieves the best known solution of G11, and its average cut value is 98.94\% of the best known solution.
The annealing time of the proposed hardware is 0.9 ms because the number of cycles is 90,000 and the clock frequency of the hardware is 100 MHz.
The proposed hardware achieves the 29 higher best cut values and the 2 higher average cut values for the King1 problem.
\begin{table}[]
    \caption{FPGA resource utilization comparisons between the proposed hardware and the SSA hardware.}
    \label{tb:t7}
    \begin{center}
        \begin{tabular}{|c|c|c|}
            \hline
            & HA-SSA & SSA \\
            \hhline{|=|=|=|}
            LUTs & 105,294 (51.67\%) & 106,139 (52.08\%) \\
            \hline
            FFs & 13,692 (3.36\%) & 13,401 (3.29\%) \\
            \hline
            BRAMs & 356 (80\%) & 356 (80\%) \\
            \hline
        \end{tabular}
    \end{center}
\end{table}
\cref{tb:t7} shows the FPGA resource comparisons between the proposed HA-SSA hardware and the SSA hardware for G11.
The main components, the spin-gate array, in the HA-SSA and SSA hardware have the same structure because they have the same number of spins and their calculation is also the same.
Therefore, there is a small difference in the utilization of look up tables (LUTs), and the utilization of flip flops (FFs) is almost the same.
The HA-SSA algorithm uses 80\% of BRAM to increase the duration of the annealing process until memory is full.
The available memory of the Genesys 2 is 13.2 M-bits; thus, each trial (150 iterations) can be operated without any interruption because its computation process ends before the memory is full.
The computation process of the annealing hardware based on conventional SSA must be interrupted every 28 iterations because the memory becomes full.
\subsection{Comparison with a related method}
\label{sec:comp}
\begin{table}
    \caption{Performance comparisons with a related method.}
    \label{tb:t6}
    \begin{center}
        \begin{tabular}{|c||c|c|}
            \hline
            & Proposed & IPAPT \cite{iccad_pt} \\
            \hhline{|=#=|=|}
            Annealing algorithm & SC-SA & IPAPT \\
            \hline
            Benchmark problem & \multicolumn{2}{c|}{G11 (MAX-CUT)}  \\
            \hline
            Best cut values & 564 & 564 \\
            \hline
            Average cut value & 558 & 561 \\
            \hline
            Annealing time & 1.00 ms & 2.64 ms \\
            \hline
            Clock frequency & 100 MHz & 150 MHz \\
            \hline
            FPGA & Kintex-7 & Vertex-5 \\
            \hline
            LUT & 105,294 & 46,753 \\
            \hline
            FF & 13,692 & 19,797 \\
            \hline
            BRAM & 356 & N/A \\
            \hline
            Bit wise of $h_i$ and $J_{ij}$ & 4 & 2 \\
            \hline
            Range of $h_i$ and $J_{ij}$ & $\{-8, -7, ..., 6, 7\}$ & $\{-1, 0, 1\}$ \\
            \hline
            \# of connection at spin & \multicolumn{2}{c|}{4} \\
            \hline
        \end{tabular}
    \end{center}
    \vspace{-5mm}
\end{table}
Several studies implement annealing hardware on the FPGAs \cite{iccad_pt} or application-specific integrated circuits (ASICs) \cite{20k_spin, statica_512}.
For evaluation purposes, our HA-SSA hardware is compared with the annealing hardware, which is based on approximated parallel tempering (IPAPT) \cite{iccad_pt}.
The parallel tempering (PT) algorithm is a variant of the SA algorithm \cite{parallel_tempering}.
In PT, the annealing process is copied to several replicas with different temperatures.
After given annealing steps, the replicas are exchanged with each other.
By exchanging the replica, PT can simulate using samples, which are obtained at the high temperature, at the low temperature.
IPAPT \cite{iccad_pt} is the approximated PT algorithm for implementing the annealing hardware.
The IPAPT hardware solved the same MAX-CUT problem (G11), and also it was implemented on the FPGA.
In addition, the IPAPT hardware has the same topology of spin connections as of which our hardware.
Our hardware is compared with the IPAPT hardware in terms of annealing time, annealing accuracy, and hardware utilization (LUT and FF).
However, the memory usage is not compared with \cite{iccad_pt} but also \cite{20k_spin,statica_512}.
Because the memory usage of the IPAPT hardware is not discussed in \cite{iccad_pt}.
In \cite{20k_spin,statica_512}, the memory usage is discussed, however, the memory is for storing the biases ($h_i$) and weights ($J_{ij}$) not for the spin states.
The memory usage for $h_i$ and $W_{ij}$ is determined by the number of spins, and the bitwise of $h_i$ and $W_{ij}$, therefore, it can not be improved by the SA algorithm.
In \cite{iccad_pt}, the Ising model solver based on IPAPT was implemented on an FPGA (Xilinx Vertex-5).
\cref{tb:t6} summarizes the comparisons between the HA-SSA hardware and the IPATP method.
The results of IPATP and HA-SSA are obtained with 100,000 annealing steps.
Each annealing process of the IPATP is repeated 1,000 times ,and HA-SSA repeats 100 times.
The IPATP approach achieves the best known solution of G11, and its average cut value is 561, with a 2.64 ms annealing time.
The proposed hardware achieves 558 as the average cut value in 1.0 ms, making it 2.64-times faster than the IPATP method.\
Our hardware utilizes 105,294 LUTs, which is 58,341 more than the IPAPT hardware.
However, the proposed hardware can represent the integer biases and weights which are in the range of -8 to 7, because the bitwise of the biases and weights is 4 bits.
In contrast, the IPAPT hardware can only represent the biases and weights which are in the range of -1 to 1.
If the biases and weights of our hardware are represented by the same bit wise as the IPAPT hardware, our hardware utilizes 88,411 LUTs.
In terms of FFs utilization, our hardware utilizes 6,105 fewer FFs than that of the IPAPT hardware.
\section{Discussion}
\label{sec:diss}
\subsection{Comparisons with SA using equivalent temperature}
\begin{figure}
    \begin{center}
        \includegraphics[width=0.9\linewidth]{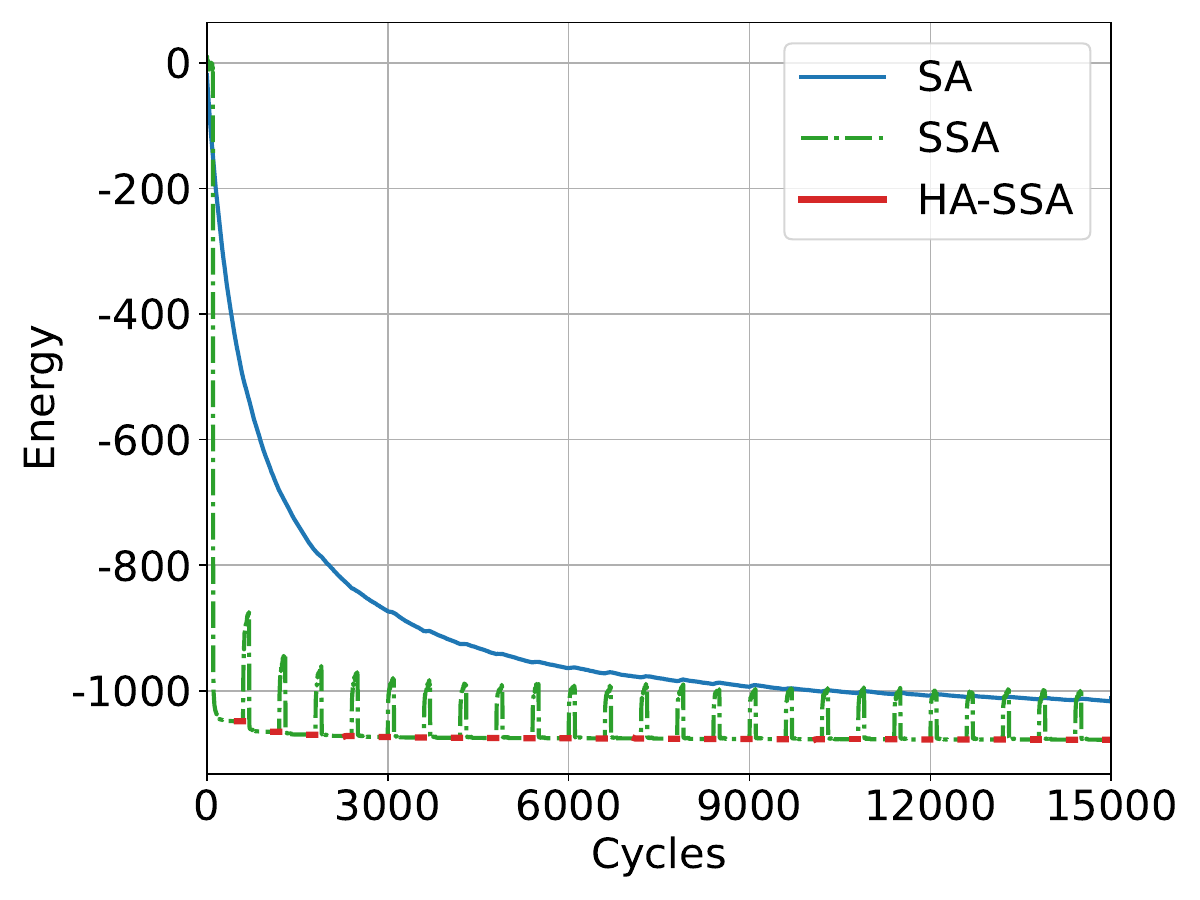}
        \caption{Ising energy comparisons during 15,000 cycles for G11 with same temperature controls.}
        \label{fig:f13}
    \end{center}
\end{figure}
The temperature of SA gradually decreases from its initial value during the entire annealing process.
In contrast, the pseudoinverse temperature of SSA and HA-SSA increases from $I_{0min}$ to $I_{0max}$ in the short period like \cref{fig:f3}.
To compare the convergence speed of SSA and HA-SSA with the speed of SA, the Ising energies from SA, SSA, and HA-SSA with the equivalent temperature control are compared.
\cref{fig:f13} shows the Ising energy comparisons during 15,000 cycles for G11 with the same temperature control.
SSA and HA-SSA use the same hyperparameters shown in \cref{tb:t2}, thus the pseudoinverse temperature increases from 1 to 32 during 600 cycles.
However, the temperature of SA should decrease from its initial value, therefore, it starts at 1 and decreases to 1/32.
With this temperature control, SA converges the Ising energy gradually, however, it can not reach the near-optimal minimum, because the duration of the annealing process is not sufficient.
The SSA and HA-SSA algorithm already converges to the near-optimal solution within 3,000 cycles.
\subsection{Effectiveness of HA-SSA for other problems}
In \cite{tnnls}, SSA solved other combinatorial optimization problems, such as traveling salesman problems (TSP) or graph isomorphism problems (GI), and obtained near-optimal solutions for each problem.
The Ising models of the GI and TSP have integer weights and integer biases not only -1 and +1 like the MAX-CUT problem.
In addition, a large K2000 MAX-CUT problem, which is a fully-connected 2,000-spin problem, is solved using the SSA algorithm \cite{katsuki_icecs}.
SSA achieved 650 times faster annealing time than that conventional SA and obtained the best known solution of K2000 (33,337).
The SSA algorithm achieves higher quality solution of K2000 compared to state-of-the-art annealing processors \cite{statica_512,cim_2000}. 
Although these results are obtained by simulations, the HA-SSA can obtain the same results as those of SSA, because the calculation of spin state is the same as SSA, and the temperature control is same when the hyperparameters are equivalent.
It means the HA-SSA algorithm can solve other complex combinatorial optimization problems with integer weights and biases, or dense connectivity.
The temperature controls for the GI and the TSP in \cite{tnnls} are same as that of HA-SSA shown in \cref{fig:f3}.
Therefore, HA-SSA can reduce the memory usage for other problems $(\log_\beta (I_{0min} / I_{0max}) + 1)$ times according to \cref{eq:cv_mem} and (\ref{eq:pr_mem}).
In \cite{tnnls, katsuki_icecs}, the solution quality is compared to various Ising machines \cite{dwave, cim_2000}.
The results of SSA are obtained by the simulations, therefore the comparison of the HA-SSA hardware with these Ising machines will be discussed in the future. 
Additionally, the several MAX-CUT problem instances, instead of the various problems, are solved by simulation and also the HA-SSA hardware.
Because the different MAX-CUT problem instances have different distributions of weights or the number of connections at each spin, the Ising models of these instances are completely different.
It means that the various instances of the MAX-CUT problems are sufficient examples for evaluating the effectiveness of HA-SSA for the various problems.
\section{Conclusion}
\label{sec:conc}
In this paper, we demonstrate the HA-SSA algorithm for realizing fast-converging annealing hardware and implement the HA-SSA hardware.
The HA-SSA method exhibits faster convergence than the conventional SA method and has memory efficiency up to 6 times greater than that of the conventional SSA algorithm by selecting the stored results using the pseudoinverse temperature.
Additionally, the pseudoinverse temperature control process is more hardware-friendly than that of the original SSA approach, so it is easy to implement in hardware.
For a performance evaluation, the proposed HA-SSA algorithm is simulated and compared to the conventional SSA algorithm and the SA algorithm.
The proposed algorithm achieves higher-quality solutions and faster annealing times for G11, G12, and G13 than the SA algorithm.
Specifically, HA-SSA achieves 6 times more memory-efficient than the conventional SSA algorithm without the loss of solution quality.
In addition, our HA-SSA hardware achieves the best known solution of G11 with a 1.00 ms annealing times, which is 2.64-times faster than the IPAPT hardware implemented on an FPGA.
The proposed HA-SSA hardware solves a MAX-CUT problem in which the edge weights are only `-1' and `+1', and its spin-gate connectivity is sparse.
In future research, other combinatorial optimization problems that have more complexity, such as problems that have integer weights and complete connectivity, are candidates for evaluating the effectiveness of the proposed algorithm and the associated hardware.
We believe that HA-SSA has the potential to be applied to real-world optimization problems.
\section*{Acknowledgment}
This work was supported by a the Japan Society for the Promotion of Science (JSPS) Grant-in-Aid for Scientific Research (B) (grant number JP21H03404), the Japan Science and Technology Agency (JST) Core Research for Evolutionary Science and Technology (CREST) (grant number JPMJCR19K3) and Doctoral Program for World-leading Innovation \& Smart Education (WISE Program) for Artificial Intelligence (AI) Electronics, Tohoku University.
\bibliographystyle{IEEEtran}
\end{document}